\documentclass{emulateapj}
\slugcomment{Accepted for publication in ApJ Letters}
\shorttitle{Intermittent accretion-powered pulsars}
\shortauthors{Lamb et al.}

\begin{document}

\title{Origin of intermittent accretion-powered X-ray oscillations in neutron stars with millisecond spin periods}

\author{Frederick K. Lamb\altaffilmark{1,3}, Stratos Boutloukos\altaffilmark{1,2}, Sandor Van Wassenhove\altaffilmark{1}, Robert T. Chamberlain\altaffilmark{1}, Ka Ho Lo\altaffilmark{1}, and M. Coleman Miller\altaffilmark{2}}

\affil{{$^1$}Center for Theoretical Astrophysics and Department of Physics, University of Illinois at Urbana-Champaign,\\1110 West Green Street, Urbana,
IL 61801-3080, USA; fkl@illinois.edu\\
{$^2$}Department of Astronomy and Maryland Astronomy Center for Theory and Computation, University of Maryland,\\College Park, MD 20742-2421, USA}

\altaffiltext{3}{Also Department of Astronomy.}

\begin{abstract}
\noindent We have shown previously that many of the properties of persistent accretion-powered millisecond pulsars can be understood if their X-ray emitting areas are near their spin axes and move as the accretion rate and structure of the inner disk vary. Here we show that this ``nearly aligned moving spot model'' may also explain the intermittent accretion-powered pulsations that have been detected in three weakly magnetic accreting neutron stars. We show that movement of the emitting area from very close to the spin axis to $\sim\,$10$\arcdeg$ away can increase the fractional rms amplitude from $\la\,$0.5\%, which is usually undetectable with current instruments, to a few percent, which is easily detectable. The second harmonic of the spin frequency usually would not be detected, in agreement with observations. The model produces intermittently detectable oscillations for a range of emitting area sizes and beaming patterns, stellar masses and radii, and viewing directions. Intermittent oscillations are more likely in stars that are more compact. In addition to explaining the sudden appearance of accretion-powered millisecond oscillations in some neutron stars with millisecond spin periods, the model explains why accretion-powered millisecond oscillations are relatively rare and predicts that the persistent accretion-powered millisecond oscillations of other stars may become undetectable for brief intervals. It suggests why millisecond oscillations are frequently detected during the X-ray bursts of some neutron stars but not others and suggests mechanisms that could explain the occasional temporal association of intermittent accretion-powered oscillations with thermonuclear X-ray bursts.
\end{abstract}

\keywords{pulsars: general -- stars: neutron -- stars: rotation -- X-rays: bursts -- X-rays: stars}

\section{Introduction}
\label{intro}

Observations made using the \textit{Rossi X-ray Timing Explorer} (\textit{RXTE}) have led to the discovery of nine neutron stars in low-mass X-ray binary systems (\mbox{LMXBs}) that produce persistent accretion-powered X-ray oscillations with periods equal to their millisecond spin periods. These accretion-powered millisecond X-ray pulsars (\mbox{APMXPs}) are thought to be weakly magnetic stars that have been spun up by accreting angular momentum (see \citealt{lamb08a} and references therein). Recently, three other neutron stars in \mbox{LMXBs} have been found to produce accretion-powered millisecond X-ray oscillations intermittently. These \textit{intermittent} \mbox{APMXPs} are the focus of this \textit {Letter}. 

An intermittent accretion-powered X-ray oscillation was first detected in HETE~J1900.1$-$2455 \citep{gall07}. The 377-Hz oscillation disappeared about two months after it was discovered, even though the star was still bright. Oscillations at $\sim\,$376~Hz have recently been discovered in X-ray bursts from this star (\citealt{watt09}). A 442-Hz accretion-powered X-ray oscillation was serendipitously detected in a $\sim\,$500~s interval during a longer \textit{RXTE} observation of SAX~J1748.9$-$2021 \citep{gavr07}. An independent analysis of all the currently available \textit{RXTE} data on this X-ray star detected a 442-Hz oscillation in 11 intervals during its 2001 and 2005 outbursts \citep{alta08, patr09}. The oscillation appeared and disappeared on a timescale of several hundred seconds. A search of the 1.3~Ms of \textit{RXTE} data available on Aql~X-1 discovered an accretion-powered X-ray oscillation at its 550-Hz spin frequency for 150~s during the peak of its 1998 outburst \citep{case08}. The spin frequency of Aql~X-1 was previously known from its X-ray burst oscillations \citep{zhan98}. Due to the poor temporal coverage of most \mbox{LMXBs} and the sparseness of the intermittent oscillations discovered so far, the total number of intermittent \mbox{APMXPs} in the Galaxy is unknown.

The accretion-powered oscillations of the known intermittent \mbox{APMXPs} share several common characteristics. All have nearly sinusoidal waveforms and fractional rms amplitudes $\la\,$3\%. The second harmonic (first overtone) of the spin frequency was rarely detected in HETE~J1900.1$-$2455; when it was, its amplitude was $\la\,$0.4\% (\citealt{gall07}). No second harmonic has been detected in the other two intermittent \mbox{APMXPs}, with upper limits \mbox{$\sim\,$0.4\%--0.9\%} (\citealt{patr09, case08}). The amplitudes of the accretion-powered oscillations produced by HETE~J1900.1$-$2455 and SAX~J1748.9$-$2021 seem to increase shortly before or after their X-ray bursts (\citealt{gall07, gavr07, alta08}).

\begin{figure*}[t!]
\hspace{0pt}
\includegraphics[scale=0.33]{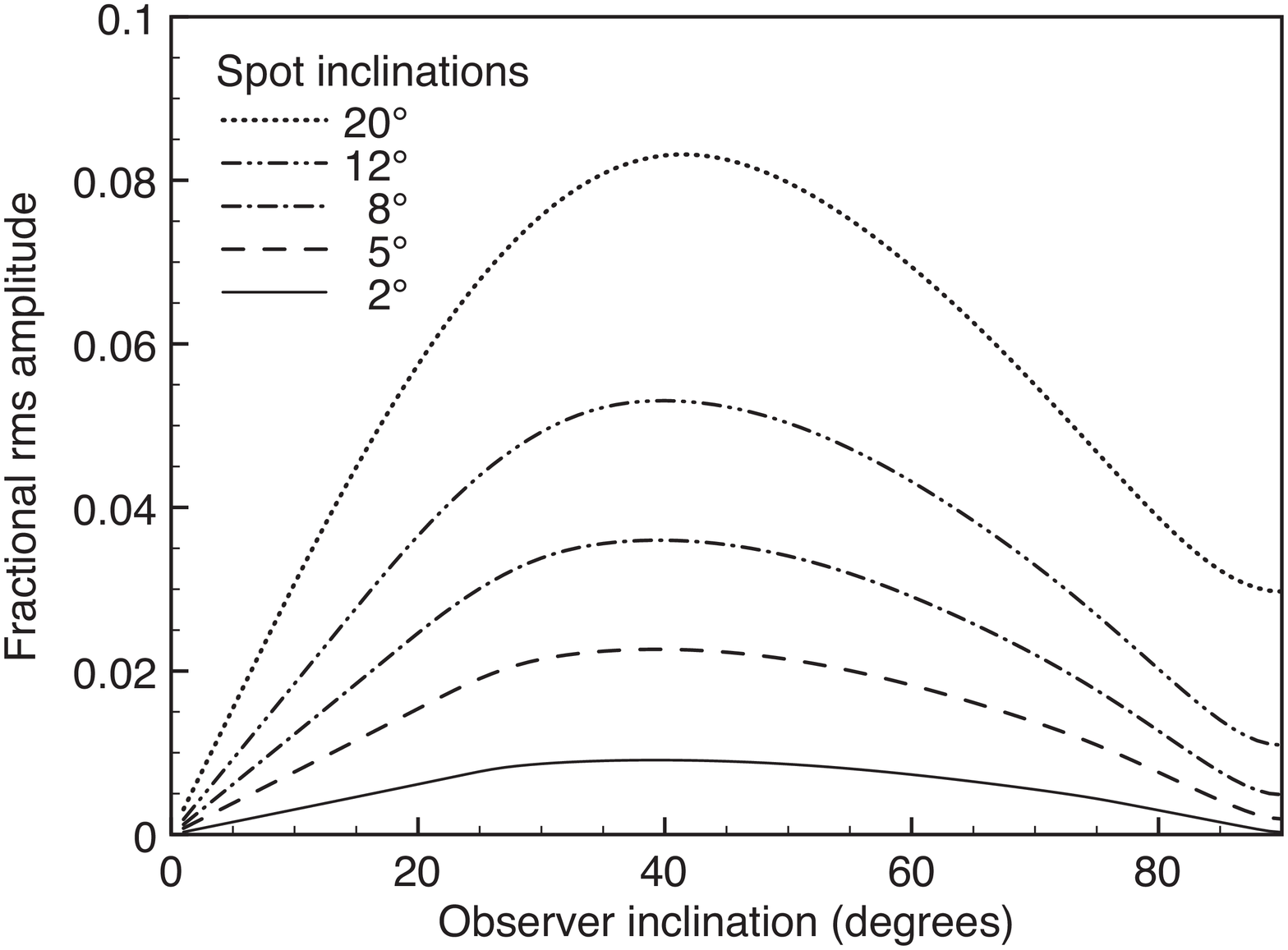}
\hspace{6pt}
\includegraphics[scale=0.33]{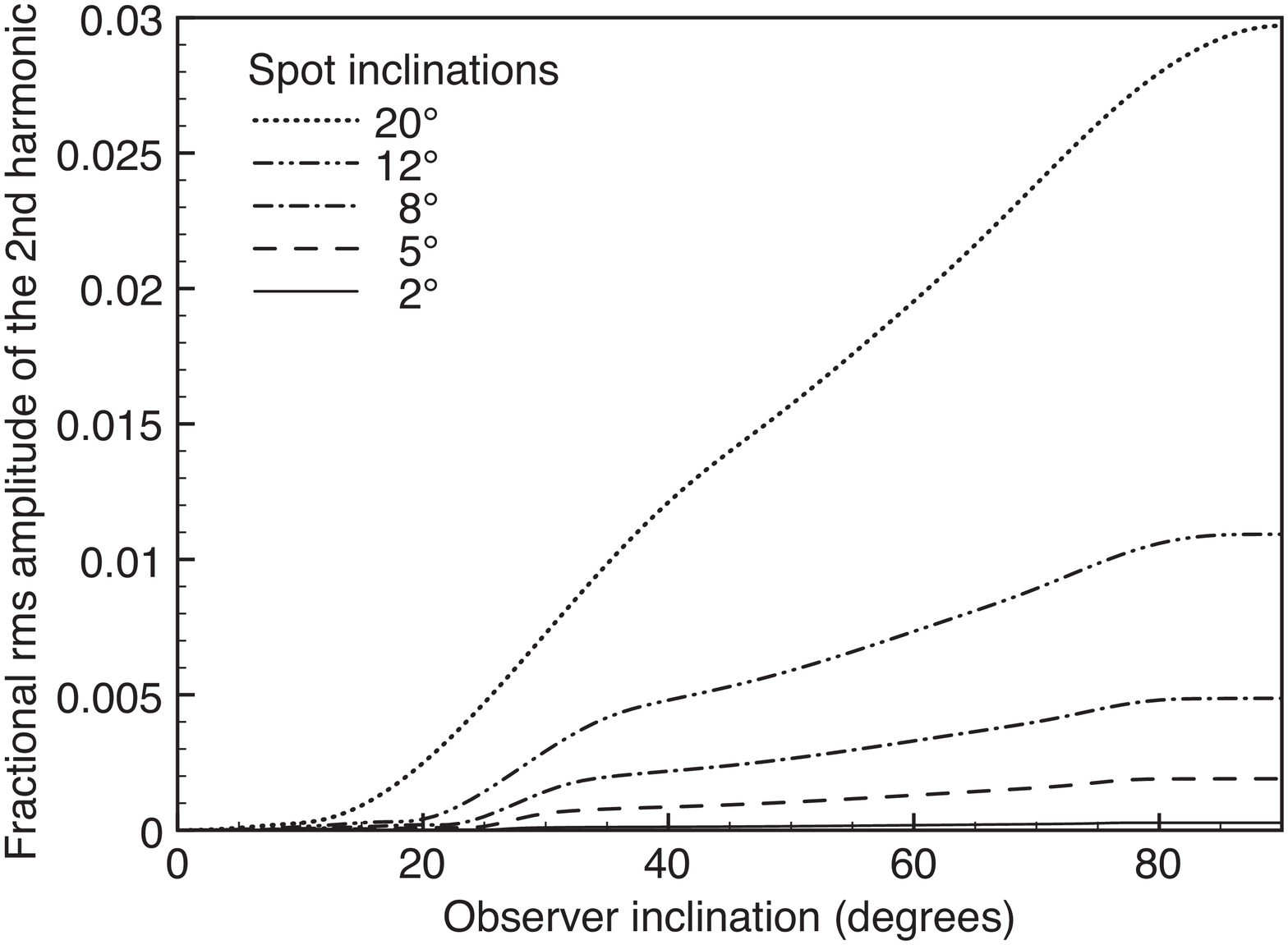}
\vspace{-6pt}
\caption{Fractional rms amplitudes of the full bolometric waveform (left) and its second harmonic component (right) as a function of the observer's inclination, for the spot inclinations shown. These waveforms are for two stable, isotropically emitting, 25$\arcdeg$-radius antipodal spots on the surface of a $1.4M_\odot$ star with a radius of $5\,GM/c^2$, spinning at 400~Hz.}
\label{fig:fractional-amplitudes}
\end{figure*}

The nearly aligned moving spot model was originally proposed to explain the  properties of the persistent \mbox{APMXPs} (see \citealt{lamb06, lamb07b, lamb08b}; \citealt{lamb09}, hereafter P1). In this model, the X-ray emitting areas are close to the spin axis but move around as the accretion rate and structure of the inner disk vary. As explained in P1, the model also has implications for the nuclear-powered oscillations observed in most of these pulsars during their thermonuclear X-ray bursts. Here we show that the model can also explain the properties of the intermittent \mbox{APMXPs}. The reason is that small changes in the latitude of the emitting area can cause oscillations to appear and disappear if the emitting area is very close to the star's spin axis (see also \citealt{lamb08b, boutloukos08}). In contrast to P1, which presented results for a wide range of spot inclinations, this paper focuses on spot inclinations $\la\,$15$\arcdeg$.

\section{Modeling and Results}
\label{model-results}

It is not yet possible to compute the accretion flows and X-ray emission of accreting millisecond pulsars from first principles, so simplified models must be used, but even these models typically have a dozen or more parameters. Despite this complexity, it is possible to identify trends and determine whether the parameter values needed to explain the observations are plausible. This is the approach we follow here. Our findings are based on the results of several hundred million waveform computations. This survey greatly extends previous work. To summarize these findings concisely, we present examples, describe trends, and discuss the parameter ranges consistent with the observations.

To illustrate our findings, we discuss the bolometric waveforms produced by isotropic or Comptonized emission like that proposed by \citet{pout03} from a single circular spot, two antipodal spots, or two spots in the same rotational hemisphere on opposite sides of the spin axis, as may occur if the stellar magnetic field evolves as described by \citet{chen98}. These models represent the emission from the stellar surface after it has been averaged over the time needed to construct a pulse profile, which is typically $\sim\,$$10^5$--$10^6$ times the $\sim\,$1~ms dynamical timescale near the neutron star.

For conciseness, we feature the representative case of spots with radii of 25$\arcdeg$ on the surface of a $1.4\,M_\odot$ star with a radius of $5GM/c^2$ (10.3~km) spinning at $400$~Hz, but we also describe how the results vary when these parameters are changed. We assume radiation from the stellar surface reaches the observer without interacting with any intervening matter.

We used the ray-tracing code described in P1 to calculate the waveforms produced by emitting spots at various rotational latitudes when viewed at different inclinations. This code uses the Schwarzschild plus Doppler (S+D) approximation \citep{mill98a}, which treats the special relativistic Doppler effects (such as aberrations and energy shifts) associated with the rotational motion of the stellar surface exactly, but treats the star as spherical and uses the Schwarzschild spacetime to compute the general relativistic redshift, trace the propagation of light from the stellar surface to the observer, and calculate light travel-time effects. The code has been validated by comparing its results with analytical results for special cases and with previously reported numerical results (see P1). For the stellar models we consider, rotational distortion of the star and frame dragging can be neglected~\citep{cade07}.

As discussed in P1, pulse amplitudes are continually as small as observed only if the emitting areas are near the spin axis, so we concentrate on this geometry. Guided by the observations cited in Section~\ref{intro}, we assume that pulsing is detectable if its fractional rms amplitude exceeds 0.5\% and that the second harmonic (first overtone) component of the pulse is detectable if its amplitude exceeds 0.5\%.

Figure~\ref{fig:fractional-amplitudes} shows some typical features of the oscillations produced by spots located within 20$\arcdeg$ of the spin axis (see also Figures~1 and~4 of P1). The fractional amplitudes are small ($\la\,$8\% in these examples) and the second harmonic is much smaller than the first, except for observer inclinations near 90$\arcdeg$. The amplitude of the first harmonic increases linearly with spot inclination for inclinations this small; it also increases linearly with observer inclination up to inclinations $\sim\,$20$\arcdeg$, but then flattens and decreases for observer inclinations $\ga\,$20$\arcdeg$. The amplitude of the second harmonic increases monotonically as the spot and observer inclinations increase.

\textit{Effect of spot geometry.} Insights into the effect of the spot geometry on the detectability of oscillations can be obtained by comparing the waveforms produced by isotropic emission from a single spot, two antipodal spots, and two spots in the same rotational hemisphere (spots with the last geometry are expected if neutron vortex motion drives both of the star's magnetic poles toward the same rotation pole; see \citealt{chen98}).

The oscillation produced by a single spot centered within 1$\arcdeg$ of the spin axis is undetectable by observers with inclinations $\la\,$35$\arcdeg$ but becomes easily detectable by almost all observers if the spot moves $\ga\,$4$\arcdeg$ toward the spin equator; the second harmonic is undetectable by all observers for spot inclinations $\la\,$8$\arcdeg$ (see Figure~1 of P1). 

The oscillation produced by antipodal spots is undetectable for any observer if the spots are within 1$\arcdeg$ of the spin axis but becomes detectable by almost all observers if the spots are $\sim\,$12$\arcdeg$ from the spin axis (see Figure~1). For this inclination, almost all observers will see amplitudes $\ga\,$2\% while observers with inclinations of 20$\arcdeg$--70$\arcdeg$ will see amplitudes of 3\%--5\%. For this spot inclination, the second harmonic is $\la\,$0.4\% and hence undetectable for observers with inclinations $\la\,$40$\arcdeg$ (again see Figure~1). 

The oscillation produced by two spots near the same rotation pole can also change from being undetectable to being easily detectable if the spots move a short distance. For example, spots at the same latitude but separated in longitude by 160$\arcdeg$ will produce oscillations that are undetectable by all observers if they are within $\sim\,$1$\arcdeg$ of the spin axis but will produce oscillations that are easily detectable (amplitudes $\sim\,$2\%--4\%) by most observers (those with inclinations $\ga\,$60$\arcdeg$) if they move $\sim\,$10$\arcdeg$ further from the spin axis. Their oscillations will also be highly sinusoidal (second harmonic amplitudes $\la\,$1\% for all observers and less than 0.5\% for observers with inclinations $\la\,$65\arcdeg).

These results illustrate two important general points. First, oscillations can change from being undetectable to being easily detectable if the emitting area is near the spin axis and moves a small distance away from it. Second, emission from two spots near the spin axis can produce highly sinusoidal oscillations even if both are visible. These conclusions are not significantly affected by variations in the spot size and beaming pattern or the stellar mass, compactness, and spin rate, for the expected ranges of these properties, as we now illustrate (see also Figures~2 and~3 of P1).

\textit{Effect of spot size.} As discussed in P1, the dependence of the fractional modulation on the spot size is relatively weak for spots with radii $\la\,$45$\arcdeg$. The reason for this is that the width of the radiation pattern produced by spots this small is determined primarily by gravitational light deflection and the beaming pattern from the stellar surface. For example, for an observer at an inclination of 60$\arcdeg$, the fractional amplitude of the waveform produced by two 5$\arcdeg$ antipodal spots is only 1.7 times greater than the amplitude produced by two 45$\arcdeg$ spots. The dependence of the fractional modulation on the spot size is even weaker for single spots, increasing by at most 10\% as the spot radius decreases from 45$\arcdeg$ to $5\arcdeg$.

\textit{Effect of stellar compactness.} As also discussed in P1, stars that are more compact tend to produce a lower fractional modulation, because the stronger light deflection averages the flux over a larger fraction of the stellar surface, reducing the observed flux variation. For example, a $1.4\,M_\odot$ star spinning at 400~Hz with two 25$\arcdeg$ antipodal spots 8$\arcdeg$ from the spin axis will produce a 5\% modulation when viewed at an inclination of 60$\arcdeg$ if its radius is $5\,GM/c^2$ but a 1.5\% modulation if its radius is $4\,GM/c^2$. As another example, a star with the values of the other parameters listed above but a radius of 10.5~km will produce a 4.2\% modulation if its mass is $1.2~M_\odot$ but only a 1.6\% modulation if its mass is $1.8~M_\odot$. However, this reduction is not in itself sufficient to explain the rare detection of accretion-powered oscillations in neutron stars in \mbox{LMXBs}, because the lowest detectable modulation is much lower than these values. Compactness is not a plausible explanation of intermittency because it changes only very slowly.

\textit{Effect of spin.} Stars that are spinning more rapidly tend to produce stronger oscillations because their higher surface velocities cause larger Doppler boosts and aberrations, making the radiation pattern more asymmetric. Fractional amplitudes also tend to increase more steeply with spot inclination for stars that are spinning more rapidly. The shapes of these curves depend on the spin rate and whether the observer sees the spot occulted by the star during its rotation. For example, an observer at 60$\arcdeg$ sees the fractional modulation from two antipodal spots inclined at 8$\arcdeg$ increase nearly linearly from 1.0\% at 100~Hz to 4.2\% at 600~Hz, whereas an observer at 25$\arcdeg$ sees the fractional modulation from the same two spots increase from 2.5\% to 3.5\% in a nonlinear fashion, because the spots are occulted for the latter observer. The fractional modulation from one or two spots increases fairly linearly for spot inclinations $\la\,$10$\arcdeg$, with a slope that increases with the spin rate. (See \citealt{pout06} for an analytical approximation valid when the emitting area is not occulted.)

\textit{Effect of the beaming pattern.} The radiation beaming pattern can affect the amplitude and harmonic content of oscillations more than the spot size or stellar compactness and spin rate, but these effects do not change the basic conclusions. To illustrate this, we discuss here the waveforms produced by the fan-like Comptonized emission beaming pattern proposed by \citet{pout03}, which differ somewhat from the waveforms produced by isotropic emission. We again consider a single spot, two antipodal spots, and two spots in the same rotation hemisphere.

The oscillation produced by Comptonized emission from a single spot is undetectable by all observers with inclinations less than 70$\arcdeg$ if the spot is within 1$\arcdeg$ of the spin axis. If the spot moves to a position 12$\arcdeg$ from the spin axis, the oscillation amplitude increases to 4\%, making it easily detectable.

The modulation produced by Comptonized emission from two antipodal spots can be up to twice as large as that produced by isotropic emission from the same spots. As before, movement of the emitting area in latitude by a few degrees will cause the oscillation to change from being undetectable to being detectable with an amplitude of a few percent, for most viewing directions.

The oscillation produced by Comptonized emission from two spots equidistant from the spin axis and 160$\arcdeg$ apart in azimuth is undetectable by all observers if they are both within 2$\arcdeg$ of the spin axis and is undetectable by observers with inclinations less than 45$\arcdeg$ if they are within 8$\arcdeg$ of the spin axis. Spots with this beaming pattern can produce modulation fractions as large as the largest observed in the  \mbox{APMXPs} when they are $\ga\,$12$\arcdeg$ from the spin axis.

The amplitudes and harmonic content of the waveforms produced by these emission geometries and beaming patterns are similar to those observed in the intermittent \mbox{APMXPs}. The fractional amplitude of the second harmonic is less than 1\% for spot inclinations $\le12\arcdeg$ and observer inclinations $\le70\arcdeg$, except when there are two spots near the same spin pole, in which case the amplitude of the second harmonic can be comparable to that of the fundamental.

\textit{Effect of rapid spot movements.} In a more realistic model that includes the expected movement of the emitting area on the surface of the star on timescales as short as the $\sim\,$1~ms dynamical time there, the range of inclinations that produce undetectable oscillations is likely to be larger than the range discussed above. The reason is that when the emitting area is near the spin axis, its azimuthal motion tends to produce large phase variations that, after averaging by the waveform reconstruction process, will reduce its apparent amplitude (see P1). This effect may help to explain the difficulty of detecting accretion-powered oscillations in accreting neutron stars with millisecond spin periods. It also means that the emitting area may have to move further from the spin axis for oscillations to become detectable.

As the emitting area moves away from the spin axis, the emission pattern becomes less axisymmetric and fluctuations in its azimuthal position produce smaller phase variations. Both effects increase the fractional modulation.

Motion of the emitting area on timescales shorter than the intervals of data used to construct a pulse profile will appear as excess background noise (\citealt{lamb85}; P1). In the physical picture proposed here, the strength of this excess noise should be correlated with the amplitude and harmonic content of the pulse profile (see P1). Motion of the emitting area on timescales longer than the intervals of data used to construct a pulse profile will produce correlated changes in the amplitude and harmonic content of the pulse profile that can be measured (again see P1).

\section{Discussion}
\label{discussion}

As explained in P1, the magnetic poles of accreting neutron stars are expected to move close to the spin axis as the stars spin up to millisecond periods. The resulting magnetic field will channel accreting gas toward the axis, causing the X-ray emitting area to be close to the spin axis. In this \textit {Letter} we have shown that if the emitting area is centered within a few degrees of the spin axis, the observed flux modulation at the spin frequency will often be so small for a wide range of viewing angles that it will be undetectable using current instruments. Rapid fluctuations in the position of the emitting area will contribute to making the oscillation at the spin frequency undetectable. These effects may explain why accretion-powered oscillations have have been detected in so few accreting neutron stars with millisecond spin periods (see also P1).

We have shown further that if the emitting area wanders toward the spin equator by $\sim\,$10$\arcdeg$, the oscillation can change from being undetectable to being easily detectable, with amplitudes comparable to those observed in the intermittent \mbox{APMXPs}. In most cases, the amplitudes of the second and higher harmonics would be detectable only when the amplitude of the fundamental exceeds several percent.

Our computations show that this behavior occurs for a variety of spot geometries and radiation beaming patterns, and a wide range of spot radii, stellar masses and radii, and observer inclinations. Infrequent episodes of detectable accretion-powered oscillations at the spin frequency are more likely for stars that are more compact, because such oscillations will generally have lower amplitudes in these stars and are therefore less likely to be detectable.

In addition to providing a possible explanation for the sudden appearance of accretion-powered oscillations at the spin frequency, our model predicts that otherwise persistent oscillations may occasionally disappear if the center of the emitting area moves too close to the spin axis. Such \textit {oscillation dropouts}, which could occur on a variety of timescales, have not yet been reported, but may be present in the existing data on known \mbox{APMXPs} or found in data on \mbox{APMXPs} discovered in the future. We encourage searches for such dropouts. 

The basic physical picture of accreting millisecond X-ray pulsars proposed here and in P1 may also explain the observed pattern of millisecond oscillation detections in the X-ray bursts of different types of neutron stars. 

Assume for the sake of argument that the magnetic poles of most weak-field accreting neutron stars in LMXBs are so closely aligned with their spin axes and the accretion flow so symmetric around the rotation pole that accretion-powered X-ray oscillations are undetectable (see P1). If the surface magnetic field is strong enough to create a symmetric pattern of nuclear burning around the spin axis (again see P1), X-ray burst oscillations will also be undetectable.

In P1 we conjectured that the magnetic fields of the persistent APMXPs SAX~J1808.4$-$3658 and XTE~J1814$-$338 are slightly asymmetric and strong enough to confine nuclear fuel, causing their nuclear burst oscillations to be locked or nearly locked to their accretion-powered oscillations, whereas the magnetic fields of other persistent APMXPs are too weak to force this.

The model of the intermittent APMXPs proposed here also suggests why accretion-powered oscillations and X-ray bursts may be associated in time, as appears to be the case in SAX~J1748.9$-$2021 \citep{alta08, patr09}. The appearance of accretion-powered oscillations signals a shift in the impact point of the accreting matter. On the one hand, such a shift could be due to a change in the accretion flow through the inner disk. This is likely to change the thermal structure and fuel loading in the outer layers of the neutron star, which might trigger a nuclear burst. On the other hand, a nuclear burst is expected to temporarily change the accretion flow in the inner disk (see, e.g., \citealt{mill96} and references therein), changing the impact point of the accreting matter and thereby possibly causing accretion-powered oscillations to become detectable.

\acknowledgments
We acknowledge helpful discussions with Diego Altamirano, Duncan Galloway, Michael Muno, Alessandro Patruno, and Michiel van der Klis. We thank Anthony Chan and Aaron Hanks for assistance in analyzing and plotting our results. These results are based on research supported by NASA grant NAG~5-12030, NSF grant AST0709015, and funds of the Fortner
Endowed Chair at Illinois, and by NSF grant AST0708424 at Maryland.

\clearpage

\end{document}